%% file: integral_sxt.tex
          [1999/12/02 1.01 (PWD)]
\documentclass[a4paper,twocolumn]{esapub} 

\input{symbols}

\input{psfig}				%

\usepackage{natbib}

\title{Multiwavelength observations revealing 
the outbursts of the two soft X-ray transients
XTE J1859+226 and XTE J1118+480}
\author[1]{S.~Chaty}
\author[1]{C.A.~Haswell}
\author[2]{R.I.~Hynes}
\author[3]{C.R.~Shrader}
\author[4]{W.~Cui}
\author[5]{C.W.~Mauche}
\author[1]{A.J.~Norton}
\author[6]{\\ J.-E.~Solheim}
\author[6]{R.~\O stensen}
\author[7]{T.R.~Geballe}
\author[8]{Z.~Ioannou}
\author[9]{A.R.~King}
\affil[1]{Department of Physics and Astronomy, The Open University, 
Milton Keynes,
UK, s.chaty@open.ac.uk}
\affil[2]{Department of Physics and Astronomy, University of Southampton,
Southampton, UK}
\affil[3]{Laboratory for High-Energy Astrophysics, 
NASA Goddard Space Flight Center, Greenbelt, USA}
\affil[4]{Department of
Physics, Purdue University, West Lafayette, IN 47907, USA}
\affil[5]{Lawrence Livermore National Laboratory, L-43, 7000 East Avenue,
Livermore, CA 94550, USA}
\affil[6]{Astrophysics Group, Department of Physics,  
University of Troms\o, Troms\o, Norway}
\affil[7]{Gemini Observatory, 670 N. A'ohoku Place, Hilo, HI 96720, USA}
\affil[8]{Astrophysics Department, University of Texas, Austin,
TX 78718, USA}
\affil[9]{University of Leicester, University Road, Leicester, LE1 7RH, UK}

\begin{document}


\maketitle

\begin{abstract}

We report here on multiwavelength observations of the two new soft
X-ray transients (SXTs) XTE J1859+226 and XTE J1118+480, which we
observed with {\it HST}/{\it RXTE}/UKIRT.
For XTE J1118+480 we also used {\it EUVE} since it is located 
at a very high galactic latitude and suffers from very
low extinction. The two sources exhibited 
very different behaviour. 
XTE J1859+226 seems quite normal and therefore
a good object for testing the accretion mechanisms in place during
the outbursts, XTE J1118+480 is much more unusual 
because it exhibits i) a low X-ray to optical ratio and ii) 
a strong non-thermal contribution in the
radio to optical domain, which is likely to be due to synchrotron emission.
We concentrate here on the near-infrared (NIR) and optical observations
of these two systems.

\end{abstract}

\section{Introduction}

The SXTs, also called X-ray novae, 
are a class of low mass X-ray binaries (LMXBs), including
GRO J1655--40 and GRO J0422+32. In this class of sources,
more than 70\% are thought to contain Black Holes \citep{charles:1998}.
The compact object accretes
matter through an accretion disk 
from a low-mass star via Roche lobe overflow. The history
of these sources is characterized by long periods of quiescence, 
typically lasting decades, and punctuated by very dramatic 
outbursts, visible at every wavelength, although these
sources are usually discovered in X-rays or the optical, 
and often accompanied by radio activity. 
A prototypical outburst is characterized by 
X-ray emission dominated by thermal emission from the hot inner
accretion disk, and optical emission produced by reprocessing of X-rays.
Two such sources were discovered during the last year: XTE J1859+226
and XTE J1118+480.
Thanks to our override programs with {\it RXTE}/{\it HST}/UKIRT we could get 
early multiwavelength observations of these systems,
and follow the evolution of these systems from outburst
towards their quiescence, 
to get more information on the mechanisms underlying
 their outbursts.
We will describe in Section \ref{1859} the observations 
and results on XTE J1859+226 and in Section \ref{1118}
those on XTE J1118+480.

\section{XTE J1859+226} \label{1859}

The first source, XTE J1859+226, was discovered by ASM/{\it RXTE} on 1999
October 9 \citep{wood:1999}, 
at the galactic coordinates: ($l,b$) = ($54.05 \deg$, $+8.61 \deg$).
This source exhibited a fast rise ($\sim 5$ days) and 
exponential decay ($\sim 23$ days) typical of X-ray novae (see Fig. 1). 
The optical counterpart reached 15th magnitude at its maximum
\citep{garnavich:1999} exhibiting a period of 9.15 +/-
0.05 hr which could be the orbital period \citep{garnavich:2000}. 
\citet{sanchez-fernandez:2000} observed
the quiescent counterpart of the source at a magnitude of 23
and also confirmed this weak photometric modulation.
A strong radio counterpart was detected, but no jet feature
was observed \citep{pooley:1999}. 
The compact object could be a black hole, i) because of the hard
X-ray spectrum which showed a high-soft state in October 1999, described by
a temperature of T $\sim 0.9$ keV and a power-law of photon index 
$\alpha \sim 1.5$,
and ii) because of the presence of high-frequency QPOs ($\geq 100$ Hz),
as in GRS 1915+105, GRO J1655--40, and XTE J1550--564 \citep{Cui:2000}.
The column density was determined as 
$N_{H} \sim 3-8 \times 10^{21} \cmmoinsdeux$ (\citet{markwardt:1999b},
\citet{dalfiume:1999}).
We undertook a multi-epoch multiwavelength program
of observations with {\it HST}/{\it RXTE}/UKIRT
(see lightcurve in Fig. 1).

	\subsection{Short timescale modulation}

To search for short timescale modulations, we observed
with the HiRAC camera on the Nordic Optical 
Telescope, used in a high speed photometric mode, during the first run of a
new camera operating system. This system allows a fast readout in a windowed
mode. The operating system was developed by R. \O stensen, and converts
existing CCD cameras which are equipped with the Copenhagen University
Observatory controllers into a fast photometer.
On Oct 17 we used an integration time of 5.3 s, readout time of 1.8 s and a
cycle time of 7.5 s, with the object and one reference star. The following
night we had a sample time of 30 s, and used 3 comparison stars.
We detected a faint modulation at 20--24 min (Fig. 2). 
This QPO may be the result of hydrodynamic oscillations at the $L_1$ point,
where a comparison with V404 Cyg predicts a value of $0.055 P_{orb}$
compared to the observed value of $0.042 P_{orb}$ \citep{king:1989}.
This modulation was also seen in several sets of data amounting to 
33 hr of optical data with different instruments/telescopes, 
in the NIR with UKIRT \citep{hynes:1999},
and much later in the outburst by \cite{charles:2000}.
No periodic modulation was seen in the NIR at shorter timescales.

	\subsection{Evolution of the Spectral Energy Distribution}

The early spectral energy distribution (SED)
observed by {\it HST} \& UKIRT (see Fig. 3) is well fitted
by a typical X-ray irradiated disk model ($T \propto R^{-3/7}$).
The model used was actually generated to fit the SED of GRO J0422+32
in outburst, which has an orbital period of 5.1 hr, and then
scaled to fit the new data with no other adjustment.
This suggests that XTE J1859+226 is also a relatively short 
period system (less than one day).
If the disk were heated by viscous processes instead of irradiation
we would expect intead to see $f_{\nu} \propto \nu^{1/3}$ 
(corresponding to $T\propto R^{-3/4}$).
Gratifyingly, this is seen in our last visit where the SED is
better fitted by a viscously heated accretion disk model 
with an edge temperature of $\sim 8000$ K,
suggesting evolution from an irradiation dominated to viscosity
dominated regime.

\section{XTE J1118+480} \label{1118}

The second source, XTE J1118+480, is a very unusual object, 
and therefore certainly very interesting.
Its galactic coordinates are ($l,b$) = ($157.62\deg$,$+62.32\deg$).
The X-ray object was discovered by {\it RXTE} on 2000 March 29 
\citep{remillard:2000} as a weak, slowly rising source, 
the post-analysis revealing
an outburst in January 2000, with a similar brightness (see Fig. 4).
The optical counterpart is a 13th magnitude star, coincident with a 18.8 mag
object in the DSS \citep{uemura:2000}.
Its optical spectrum was typical
of X-ray novae in outburst \citep{garcia:2000}.
This system was characterized by a very low X-ray to optical flux ratio
of 5 (see \citet{remillard:2000} \& \citet{uemura:2000}),
when the typical value is 500 (see e.g. \citet{tanaka:1996}).
A weak photometric modulation on 4.1 hr (0.17082 d) 
period was rapidly discovered \citep{cook:2000}, which was associated
with the orbital period, the shortest among the black hole
candidates. 
Flickering with an amplitude of $\sim 0.4$ mag,
and also a quasi-periodic oscillation (QPO) 
at 10 s, was observed in the optical, in the UV
\citep{haswell:2000c} and also in the X-rays, with an evolving frequency
\citep{wood:2000}.
A faint radio counterpart was detected at 6.2 mJy, but no jet feature
could be spatially resolved.
The large value of the mass function, f(M) = 5.9 +/- 0.4 solar masses,
suggests that the compact object is a black hole (\citet{wagner:2000} \&
\citet{mcclintock:2000a}).

Its location at an unusually high galactic latitude ($b=+62 \deg$)
at an estimated distance of $0.8 \kpc$
\citep{mcclintock:2000}
implied that there is a very low absorption along the line of sight
of the source, with a column density estimated to
$N_{H} \sim 0.75-1.3 \times 10^{20} \cmmoinsdeux$ (\citet{hynes:2000},
\citet{mcclintock:2000}).
Triggering our multi-epoch multiwavelength program
 with {\it HST}/{\it RXTE}/UKIRT and requesting Director's Discretionary
{\it EUVE} observations,
we got unprecedented broadband 
coverage (see lightcurve in Fig. \ref{1118_lc} \& SED in Fig. 6,
more details in \cite{hynes:2000}).
The SED suggests that the system was
exhibiting a low-state mini-outburst, with the inner radius
of the accretion disk at $\sim 2000 R_s$ ($R_s$: Schwarzschild
radius).
One of the most striking features 
was the strong non-thermal (likely synchrotron)
contribution
in the optical and NIR wavelengths.
Indeed, the SED shows a very flat spectrum from the NIR to the
UV ($\sim 1000  - 50\,000 \AA$),
suggesting that there is another source of NIR flux apart from thermal
disk emission, likely related to radio emission, and therefore
possibly synchrotron.

	\subsection{Non-thermal contribution}

The $\sim 10$ s QPO was also seen in the UV, implying a common
nature for this QPO from the optical to the X-rays.
We looked for a rapid periodicity in the NIR wavelengths,
and we therefore observed with the 
3.8-m U.K. Infrared Telescope and IRCAM3 instrument on 
June 24.23 UT when the source was at K = 11.512 +/-- 0.004 and 
on July 15.25 UT, when the source was at K = 11.948 +/-- 0.006.
The 0.02 mag/day mean decline between these two dates was 
therefore much stronger than from April to June
\citep{hynes:2000}, and
comparable to the 0.015 mag/day mean decline seen in the optical
(VSNET observations).
During both observations we took rapid photometric observations
for nearly 1 hour, with 2 s integration time on the source
and nearly 7 s sampling time. The source showed rapid fluctuations
through nearly 0.5 mag, with a 0.05 mag error bar for each individual
frame. A periodicity search analysis in both runs
did not show any secure quasi-periodic oscillation,
although there could be some modulations around 35 s.
Unfortunately our sampling time could not address the presence
or absence of the periodicity of 10 s seen in the optical and in
the X-rays.

However, we could detect flickering at NIR wavelengths, of bigger
amplitude ($\sim 0.8$ mag) than in the optical ($\sim 0.4$ mag).
This flickering in the K band is reported in Figure 5.
We also took a NIR K-band spectrum of this source using the
CGS 4 instrument and a $0.6 \asec$ slit
on June, 27.2 UT, which was featureless, also consistent with the fact
that the disk is not the only source of emission in this part of the spectrum.
These two facts suggest a strong non-thermal (likely synchrotron) emission in
the NIR wavelengths.
This is consistent with the flat slope of 
the SED, shown in Figure 6
from the radio to the IR, and will be developed in Chaty et al. (in prep.).

\subsection{The nature of the system}

In our {\it HST} UV spectra we observed a complete absence
of carbon and oxygen lines, suggesting that the material
accreted from the companion star has undergone significant
CNO-cycle processing resulting in C and O depletion \citep{haswell:2000b}.
Since a 4 hr binary would normally be thought to contain an M dwarf
companion, no CNO processing would be expected.
Therefore it is likely that the companion star is the core of
a larger star that has lost its envelope: a stripped giant.
To test this hypothesis and try to reveal the nature 
of the companion star of the system, we will need to observe this source
in quiescence with spectroscopic
observations, at both optical and NIR wavelengths.

\section{Conclusions}

XTE J1859+226 allows us to study the accretion mechanisms 
during the outburst and the changes in temperature distribution
due to irradiation.
XTE J1118+480 allowed us to get
broadband coverage from the radio to the X-rays, showing 
that it was exhibiting a very low-state mini-outburst,
with a strong non-thermal (likely synchrotron) contribution.

\section*{acknowledgements} 
S.C. is very grateful to the UKIRT staff for their availability and skills
to perform service observations for override programs,
and in particular to John Davies, Sandy Leggett and Andy Adamson.
We thank {\it HST/STScI} and {\it RXTE} staff for ongoing effort in these
multi-epoch campaigns, and {\it EUVE} staff for the DD observations.
S.C., C.A.H. R.I.H. and A.R.K. acknowledge
support from grant F/00-180/A from the Leverhulme Trust.

   \bibliographystyle{aa}
   \bibliography{science}

\twocolumn
\newpage

\begin{figure}
\centerline{\psfig{file=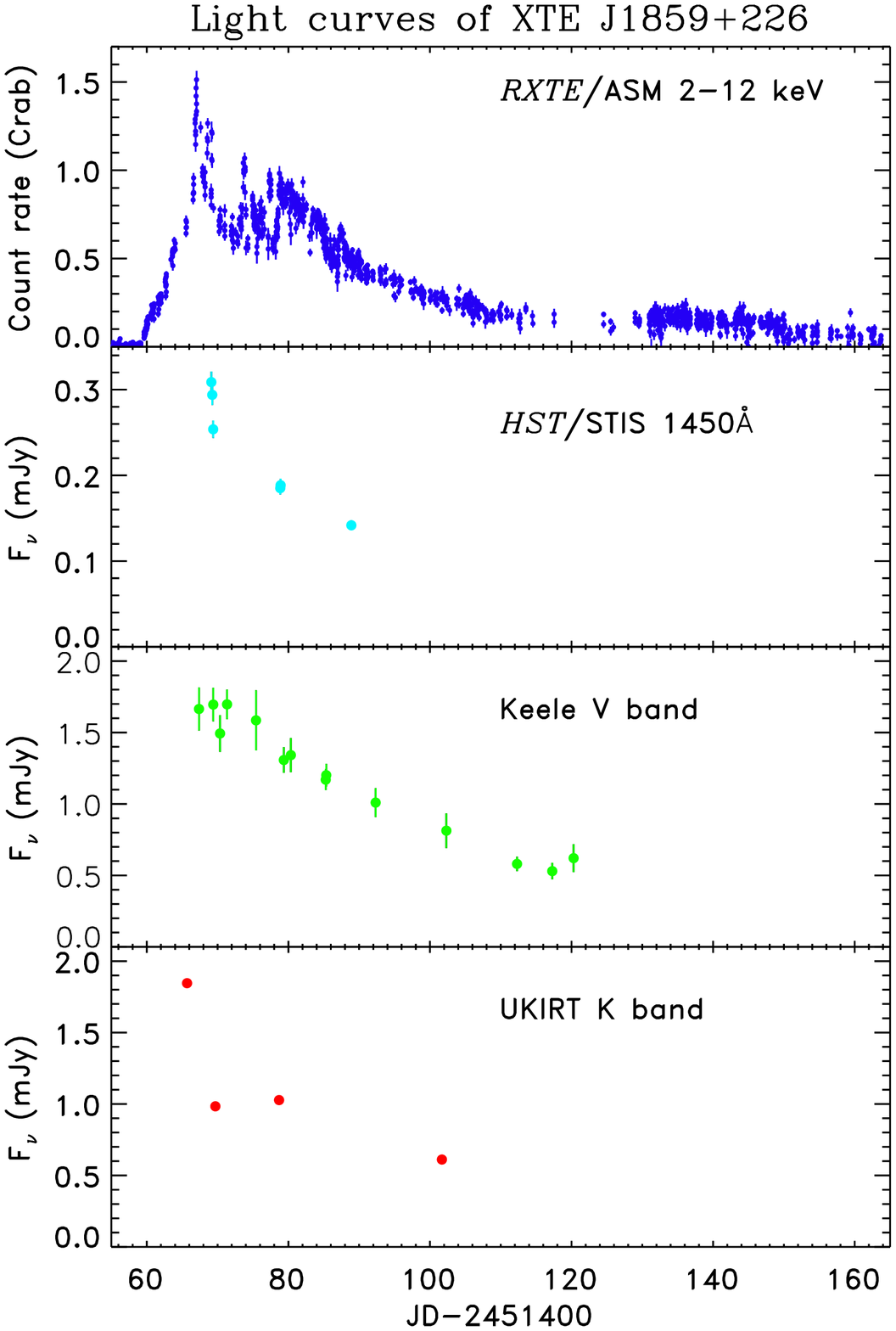,angle=0.,width=7.cm}}
{\bf Figure 1 -- The Outburst History:}
{\small \it Respectively from top to bottom are shown: {\it RXTE}/ASM 2-12 keV counts, {\it HST}/STIS 
1450 Angstr\"oms, Keele V, and UKIRT K bands.
} \label{1859_lc}
\end{figure}


\begin{figure}
\centerline{\psfig{file=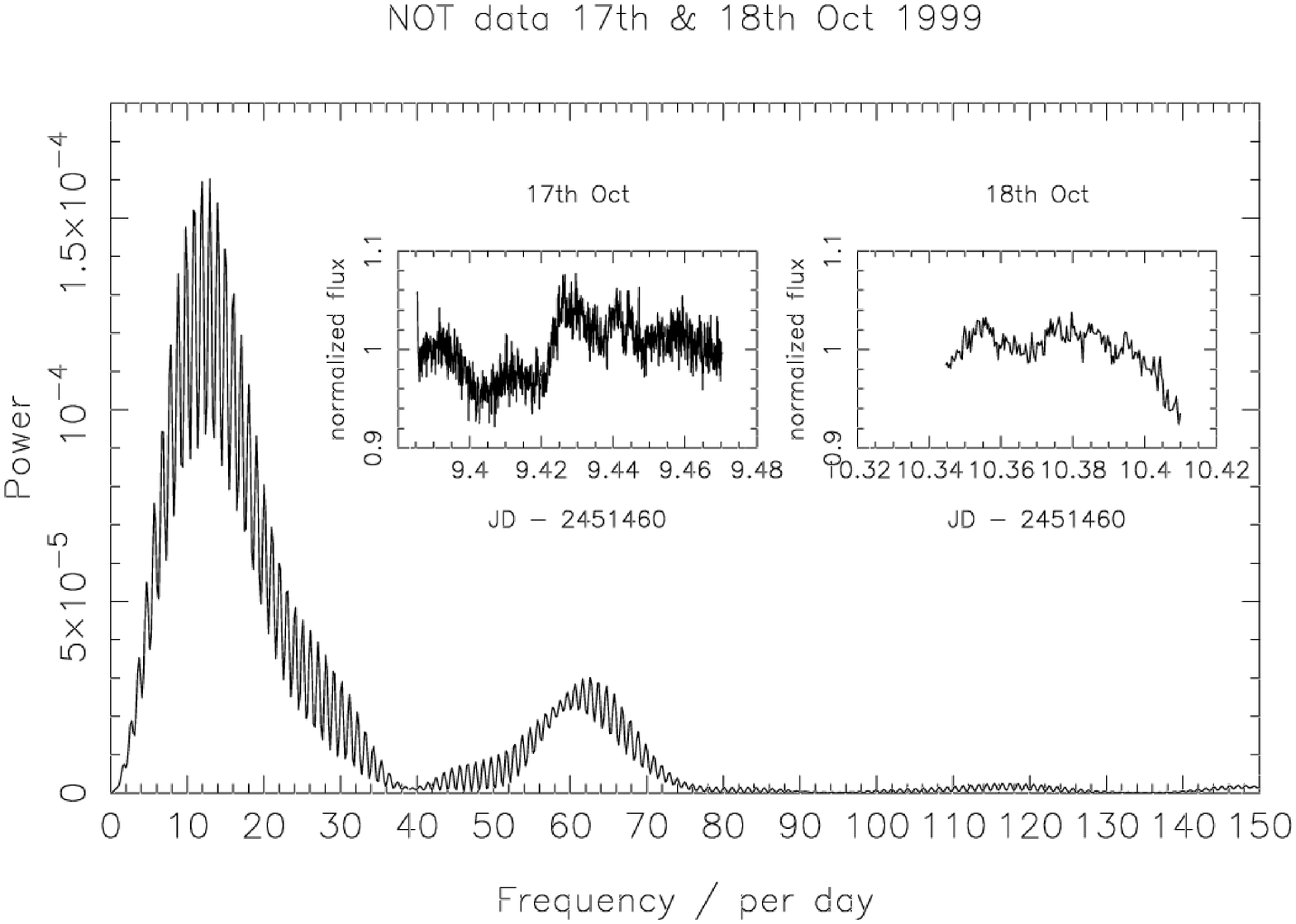,angle=-90.,width=7.cm}}
{\bf Figure 2 -- Periodicities:}
{\small \it A $\sim 23$ min periodicity (frequency $\sim 62$ per day)
was seen in the Nordic Optical Telescope
data acquired on 1999 October 17--18th, perhaps 
related to a Lagrange point oscillation.} \label{1859_not}
\end{figure}


\begin{figure}
\centerline{\psfig{file=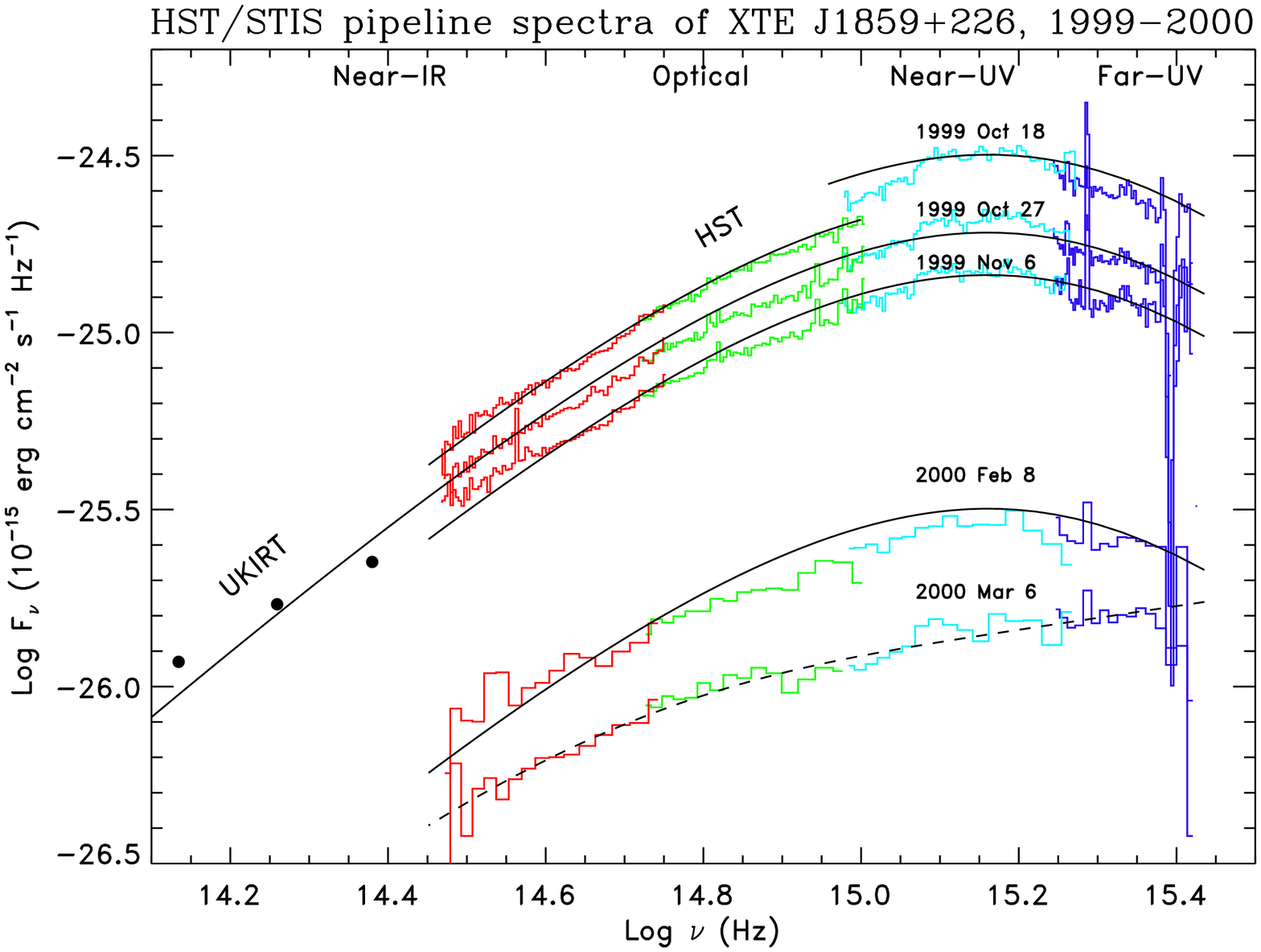,angle=+90.,width=7.cm}}
{\bf Figure 3 -- The irradiation view:}
{\small \it Note the change from an irradiated spectrum during the outburst
(top curves) to a viscously heated disk after the outburst
(bottom curve).} \label{1859_ir}
\end{figure}


\begin{figure}
\centerline{\psfig{file=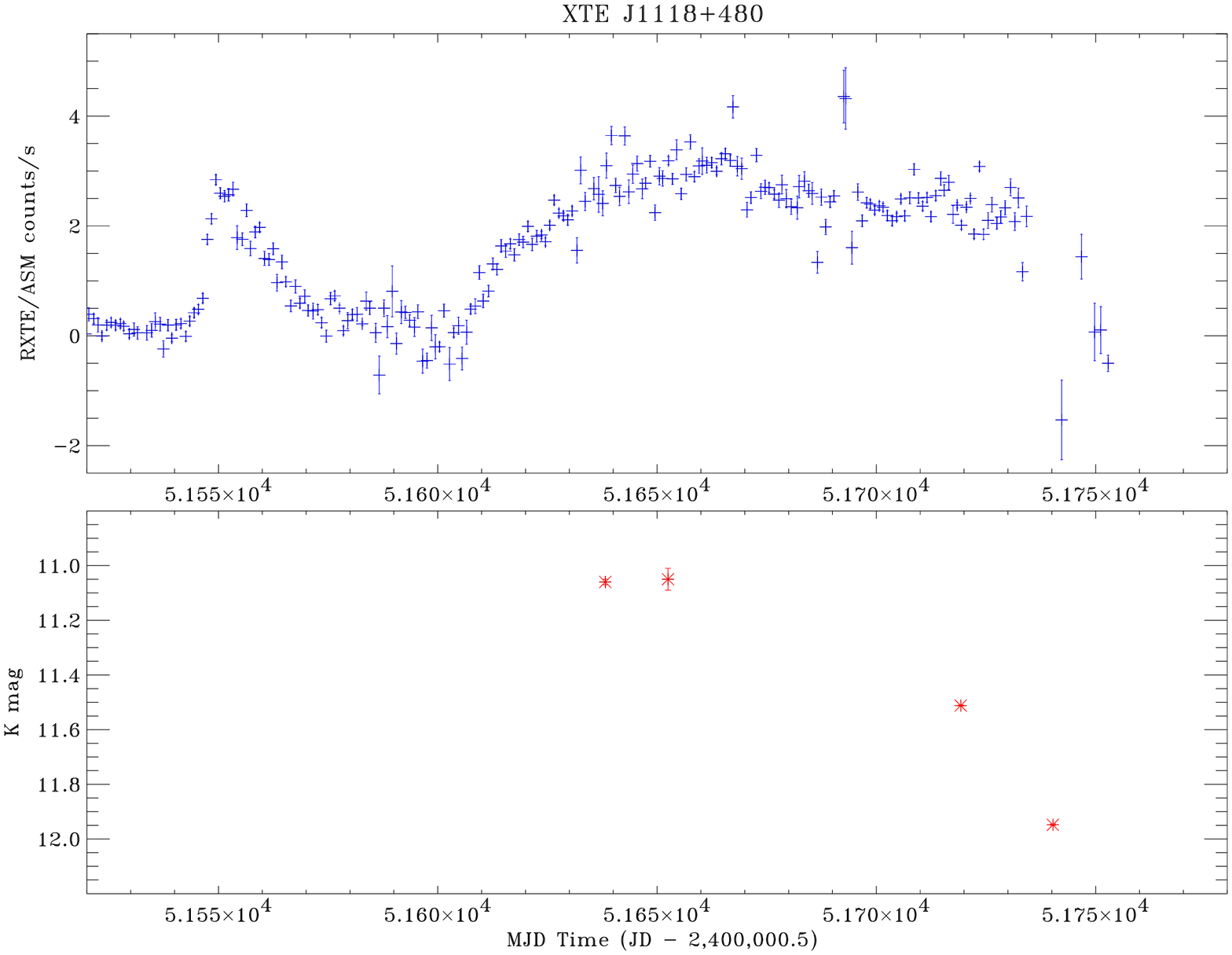,angle=90.,width=8.cm}}
{\bf Figure 4 -- The Outburst History:}
{\small \it {\it RXTE}/ASM 2-12 keV counts (top) \& UKIRT K bands (bottom).
} \label{1118_lc}
\end{figure}



\begin{figure}
\centerline{\psfig{file=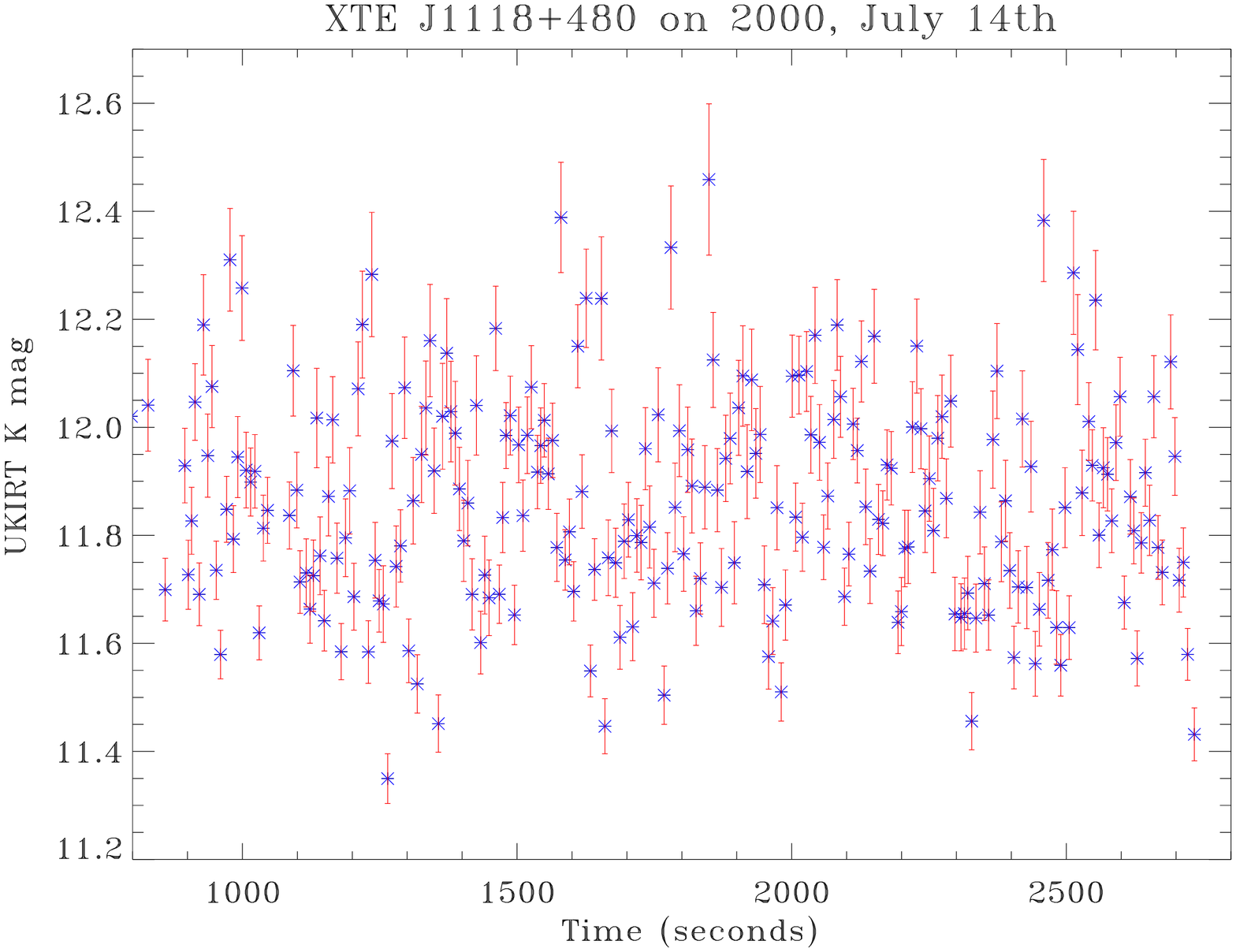,angle=90.,width=8.cm}}
{\bf Figure 5 -- IR Flickering:}
{\small \it Significant Flickering as seen in K band with UKIRT, on a 0.8 mag amplitude
with error bars of only 0.1 mag.} \label{1118_flick}
\end{figure}


\begin{figure}
\centerline{\psfig{file=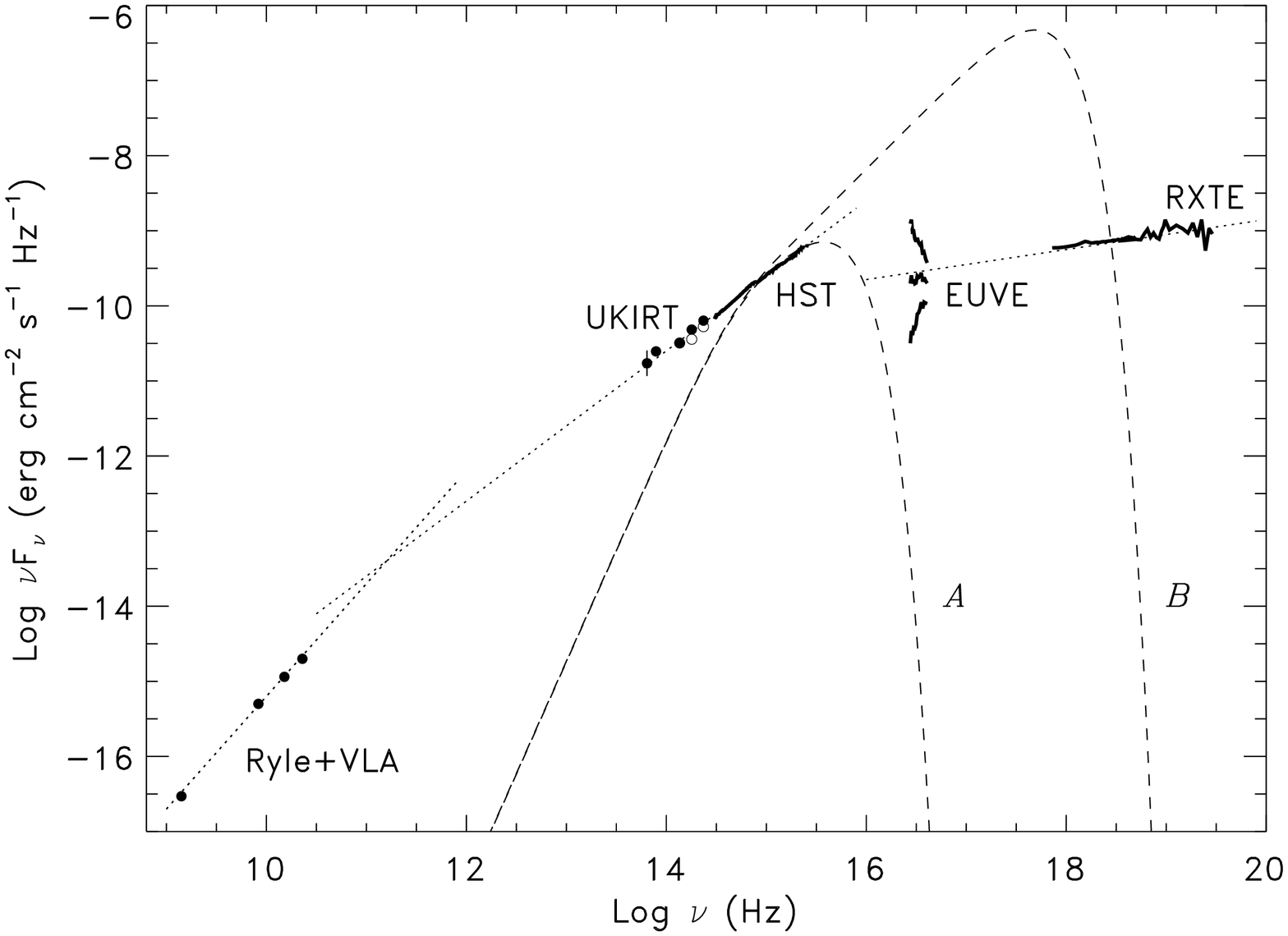,angle=90.,width=8.cm}}
{\bf Figure 6 -- \small Spectral Energy Distribution:}
{\small \it The {\it EUVE} fluxes are corrected with $N_H = 0.35, 0.75, \& 1.15 \times
10^{20} \cmmoinsdeux$. Dashed lines: two steady-state
disk models with an outer disk layer  at 8000K and internal disk
layer at 2000 Rs (model A) and 3 Rs (model B).
Dotted lines: different power laws, with spectral indices of 0.5,
0.0 and -0.8 \citep{hynes:2000}.} \label{1118_sed}
\end{figure}


\end{document}

%% file: symbols.tex
\def\cmmoinsdeux{\mbox{ cm}^{-2}}

\def\kpc{\mbox{ kpc}}

\def\deg{^{\circ}}

\def\asec{^{\prime \prime}}

\def\ltsima{\; \buildrel < \over \sim \;}
\def\simlt{\lower.5ex\hbox{\ltsima}}            
\def\gtsima{\; \buildrel > \over \sim \;}
\def\simgt{\lower.5ex\hbox{\gtsima}}            
